\documentclass{article}
\usepackage{INTERSPEECH2020,amsmath,graphicx,url,times}
\usepackage{color}
\usepackage[pdftex,pdftitle={Improving Opus Low Bit Rate Quality with Neural Speech Synthesis},pdfauthor={Jan Skoglund, Jean-Marc Valin}]{hyperref}
\usepackage{INTERSPEECH2020}

\title{Improving Opus Low Bit Rate Quality with Neural Speech Synthesis}

\name{Jan Skoglund$^1$, Jean-Marc Valin$^2$\sthanks{\hspace{2mm}This work was performed while the author was with Mozilla.}}  

\address{
  $^1$Google, San Francisco, CA, USA \\
  $^2$Amazon, Palo Alto, CA, USA}

\email{jks@google.com, jmvalin@amazon.com}

\begin{document}

%\ninept
\maketitle

\begin{sloppy}

\begin{abstract}
The voice mode of the Opus audio coder can compress wideband speech at bit rates
ranging from 6~kb/s to 40~kb/s.
However, Opus is at its core a waveform matching coder, and
as the rate drops below 10~kb/s, quality degrades quickly. As the 
rate reduces even further,
parametric coders tend to perform better than waveform coders.
In this paper we propose a backward-compatible way of improving low bit rate Opus quality
by re-synthesizing speech from the decoded parameters. We compare two different neural 
generative models, WaveNet and LPCNet. WaveNet is a powerful, high-complexity,
and high-latency architecture
that is not feasible for a practical system, yet provides a best known  
achievable quality with generative models. LPCNet is a low-complexity, 
low-latency RNN-based generative model, and practically 
implementable on mobile phones. We apply these systems with parameters from Opus 
coded at 6~kb/s as conditioning features for the generative models. A listening test 
shows that for the same 6~kb/s Opus bit stream, synthesized speech using LPCNet clearly
outperforms the output of the standard Opus decoder.
This opens up ways to improve the decoding quality of existing speech and audio waveform coders
without breaking compatibility.
\end{abstract}

\section{Introduction}
\label{sec:intro}
Speech compression methods operating at very low bit rates often represent speech as a sequence of parameters extracted at the encoder. 
These systems are referred to as parametric coders or vocoders.
Speech generated at the decoder from the transmitted parameters sounds \emph{similar} to the original speech, 
but the waveform does not match the original. The resulting speech often sounds intelligible but with a robotic character. 
Examples are linear predictive vocoders \cite{tremain:76,McCree96} or sinusoidal coders \cite{Hedelin:81,McAulay:86}. 
Another family of coders is the hybrid waveform coders, which use some signal modelling, yet try to mimic the signal waveform.
A typical example is code excited
linear prediction (CELP) \cite{schroeder1985code}. Hybrid coders are used in most mobile telephony 
and VoIP standards, examples are the AMR coder
\cite{3GPP-AMR,bessette2002adaptive} and the IETF Internet coder Opus~\cite{rfc6716}.
However, because these schemes attempt to reconstruct the signal waveform, they require higher rates to be successful, 
and at very low rates, say below 6~kb/s, the quality of 
waveform matching hybrid coders eventually becomes 
inferior even to the quality of parametric coders. 

Generative systems using neural speech synthesis have recently
demonstrated the ability to produce high quality speech.
The first method 
shown to provide excellent
speech quality was WaveNet \cite{van2016wavenet}, originally proposed for text-to-speech synthesis. 
Since then, WaveNet has been used for parametric coding that significantly out-perform
more traditional vocoders, either using an existing vocoder
bit stream~\cite{Kleijn2017wavenet}, or with a quantized learned
representation set~\cite{garbacea2019wavenet}. A typical WaveNet configuration requires a very
high algorithmic complexity, in the order of hundreds of GFLOPS, along with a
high memory usage to hold the millions of model parameters. Combined with the high latency,
in the hundreds of milliseconds, this renders WaveNet impractical
for a real-time implementation. Replacing the dilated convolutional networks 
with recurrent networks improved memory efficiency in SampleRNN~\cite{mehri2016samplernn},
which was shown to be useful for speech coding in~\cite{klejsa2019}.
WaveRNN~\cite{kalchbrenner2018efficient} also demonstrated possibilities for synthesizing
at lower complexities compared to WaveNet. Even lower complexity and real-time operation was
recently reported using LPCNet~\cite{lpcnet}.

These previously proposed systems are all based on 
quantized parametric speech coder features as conditioning to the 
neural speech generation.
In this work, we demonstrate the ability
of generative networks to improve the synthesis quality of the hybrid,
waveform-matching, Opus coder (Section~\ref{sec:opus}) operating at a low bit rate.
The goal here is to improve the quality of an existing waveform coder
without changing the bit stream. The 
results may hopefully encourage the use of neural synthesis to improve the quality
of other standard coders at low rates, since
deploying new coders to replace existing ones is a long and complicated process.
This approach can thus help 
extending the life of existing coders without introducing compatibility issues during transitions.

For this task, we consider both WaveNet and LPCNet models in Section~\ref{sec:generativet}.
Section~\ref{sec:training} describes the conditioning features and training procedure,
and we evaluate the two models in Section~\ref{sec:evaluation}. We then conclude in
Section~\ref{sec:conclusion}.

\section{Opus speech compression}
\label{sec:opus}
Opus~\cite{rfc6716} is a versatile coder that supports narrowband (8~kHz sampling frequency)
to fullband (48~kHz sampling frequency) speech and audio. 
For speech communication it has hundreds of millions of users through applications such as 
Zoom and \mbox{WebRTC}~\cite{WebRTC, rfc7478} based ones, such as
Microsoft Teams, Google Meet, Duo, Skype, and Slack. 
Opus is also one of the main audio coders used in YouTube for
streaming.

It is based on a combination of a linear predictive coding
part~(SILK~\cite{vos2013voice}) and a transform coding part~(CELT \cite{Valin2013}). 
In this work we focus on wideband
speech, i.e., a sampling frequency of 16~kHz, using the SILK-only mode of Opus.

For wideband speech, SILK uses 16$^\mathrm{th}$-order
linear prediction coefficients~(LPC).
The long-term predictor~(LTP) uses a 5-tap filter, which both controls the
amount of prediction as a function of frequency and, to some extent, provides
some of the benefits of a fractional pitch period.
The Opus reference encoder\footnote{\url{https://opus-codec.org/}} we use for this work
jointly optimizes the LPC and LTP to minimize the residual signal variance.

SILK is a linear prediction-based coder that uses noise feedback coding (NFC) \cite{makhoul1979} rather than 
regular CELP \cite{schroeder1985code}.
The residual is coded as a sum of pulses, plus a pulses-dependent dither signal.
Note that even though there is technically a separation into a spectral envelope filter 
and an excitation, both SILK and CELP coders are indeed hybrid waveform coders, with weighted
waveform matching loss functions.

Because SILK uses entropy coding, it is fundamentally also a variable bit rate~(VBR) coder. 
Rather than having a fixed bit allocation for its quantizers, it uses
rate-distortion optimization (RDO) for both the filter and the residual. The number of
bits allocated to the filter (represented as line spectral pairs \cite{itakura1975}) does not vary significantly with the
total bit rate. As a result, the number of bits used for the residual goes down rapidly as the
total bit rate decreases below 8~kb/s, with the quality eventually becoming unacceptable.
By using a neural synthesis on the Opus bit stream, it is possible to create a decoder
that degrades more gracefully without breaking compatibility.

\section{Autoregressive Generative Networks for Speech Synthesis}
\label{sec:generativet}
Autoregressive neural synthesis systems are based on the idea that the speech signal probability 
distribution $p(\mathbf{S})$ can be factorized as a product of conditional probabilities 
\cite{van2016wavenet}
\begin{equation}
p(\mathbf{S}) =\prod_{t = 1}^T p(s_t |s_{t-1}, s_{t-2}, \ldots, s_2, s_1),
\end{equation}
where $\mathbf{S}=\{s_1, s_2, \ldots, s_T\}$ is a set of
consecutive speech samples.
The probability of each speech sample $s_t$ is then conditioned on previous samples, 
i.e., a tractable \emph{scalar} autoregressive structure $p(s_t |s_{t-1}, s_{t-2}, \ldots)$.
A practical speech generation system will also need additional conditioning features,
$\boldsymbol{\theta}_t$, to
guide the waveform generation. Examples of such features are spectral envelope information, pitch, and gain. 
The output sample $s_t$ is then drawn from
the distribution $p(s_t |s_{t-1}, s_{t-2}, \ldots, \boldsymbol{\theta}_t)$, modelled through a neural net as 
$p(s_t| u(\boldsymbol{\theta}_t, s_{t-1}, s_{t-2}, \ldots, \boldsymbol{\omega}))$,
where $u$ denotes a deterministic neural 
network with parameters (e.g., weights)
$\boldsymbol{\omega}$.
Examples of distributions utilized in generative systems are discrete softmax \cite{van2016pixelrnn} and
mixtures of logistics \cite{salimans2017pixelcnn++}. 

In this work, we explore two autoregressive models for synthesizing speech from Opus
parameters. We use WaveNet as an ``informal upper bound" that demonstrates the highest obtainable
quality from generative networks. To demonstrate what can currently be achieved in real time
on general purpose hardware, we also explore LPCNet as generative synthesis.

\subsection{WaveNet}
The WaveNet architecture is a deep multi-layer structure using dilated convolution with 
gated cells. The number of layers are typically more than 25 
and the conditional variables are supplied to all layers of the network. 

A convolutional neural network has a finite memory - the receptive field - 
which depends on the number of layers in the network. 
During training, WaveNet learns the parameters of a (discretized) mixture of 
logistics function 
that represents the conditional discrete probability distribution $p(s_t)$. 
The WaveNet architecture
has shown impressive speech quality for text-to-speech \cite{van2016wavenet} and low bit rate speech coding 
\cite{Kleijn2017wavenet, garbacea2019wavenet}.
This performance comes at the price of a high complexity, typically 100+ GFLOPS,
high memory requirements with millions of network parameters, and a high latency, 400+~ms. 
Even though more recent generative architectures, such as 
WaveRNN~\cite{kalchbrenner2018efficient}, have shown the ability to 
operate at a low complexity, it comes at the cost of trading off quality.
At this low complexity the recurrent architectures have not been 
able to fully match the quality of the original WaveNet. 
We use a WaveNet model with 27 layers (9 dilation steps) and 256 hidden states 
in each layer, and with a receptive field of 192~ms. The output of the network is a 
logistic mixture distribution sampled to produce wideband speech at a 16-bit resolution.

\subsection{LPCNet}
The WaveRNN model~\cite{kalchbrenner2018efficient} is based on a
sparse gated recurrent unit (GRU)~\cite{cho2014properties} layer.
LPCNet~\cite{lpcnet} improves on \mbox{WaveRNN} by adding linear
prediction, as shown in Fig.~\ref{fig:Overview-of-LPCNet}.
Linear prediction is long known \cite{atal1971speech} to represent the spectral envelope 
of speech very well, and this enables the non-linear components of the network to
focus on a spectrally flat excitation waveform.  
LPCNet is
divided in two parts: a \emph{frame rate network} that computes conditioning
features for each frame, and a \emph{sample rate network}
that computes conditional sample probabilities. In addition to using
the previously generated speech sample $s_{t-1}$, the sample rate network also uses
the 16\textsuperscript{th} order prediction $y_{t}=\sum_{i=1}^{16}a_{i}s_{t-i}$
and the previously generated excitation $e_{t-1}$, where $e_{t}=s_{t}-y_{t}$. 

LPCNet generates speech signals at an 8-bit resolution using 
$\mu$\nobreakdash-law companding.
To shape the
quantization noise and make it less perceptible a pre-emphasis filter
$E(z)=1-\alpha z^{-1}$ is applied on the input speech (with $\alpha=0.85$)
and the inverse de-emphasis filter on the output. 
A major complexity saving comes 
from the insight that since $s_{t-1}$,
$y_{t}$, and $e_{t-1}$ are discrete, the contribution
$\mathbf{v}_{i}^{\left(\cdot,\cdot\right)}$ of each possible value
to the gates and state of $\mathrm{GRU_{A}}$ in  Fig.~\ref{fig:Overview-of-LPCNet} can be pre-computed.
In addition, the contribution $\mathbf{g}^{\left(\cdot\right)}$
of the frame rate network to $\mathrm{GRU_{A}}$ can be computed only
once per frame. After these simplifications, only the recurrent matrices
$\mathbf{W}_{\left(\cdot\right)}$ remain and the sample rate network
is then computed as (biases omitted for clarity)
\begin{align}
\mathbf{u}_{t}= & \sigma\left(\mathbf{W}_{u}\mathbf{h}_{t-1}+\mathbf{v}_{s_{t-1}}^{\left(u,s\right)}+\mathbf{v}_{y_{t}}^{\left(u,y\right)}+\mathbf{v}_{e_{t-1}}^{\left(u,e\right)}+\mathbf{g}^{\left(u\right)}\right)\nonumber \\
\mathbf{r}_{t}= & \sigma\left(\mathbf{W}_{r}\mathbf{h}_{t-1}+\mathbf{v}_{s_{t-1}}^{\left(r,s\right)}+\mathbf{v}_{y_{t}}^{\left(r,y\right)}+\mathbf{v}_{e_{t-1}}^{\left(r,e\right)}+\mathbf{g}^{\left(r\right)}\right)\label{eq:LPCNet}\\
\widetilde{\mathbf{h}}_{t}= & \tau\left(\mathbf{r}_{t}\circ\left(\mathbf{W}_{h}\mathbf{h}_{t-1}\right)+\mathbf{v}_{s_{t-1}}^{\left(h,s\right)}+\mathbf{v}_{y_{t}}^{\left(h,y\right)}+\mathbf{v}_{e_{t-1}}^{\left(h,e\right)}+\mathbf{g}^{\left(h\right)}\right)\nonumber \\
\mathbf{h}_{t}= & \mathbf{u}_{t}\circ\mathbf{h}_{t-1}+\left(1-\mathbf{u}_{t}\right)\circ\widetilde{\mathbf{h}}_{t}\nonumber \\
p\left(e_{t}\right) & =\mathrm{softmax}\left(\mathrm{dual\_fc}\left(\mathrm{GRU_{B}}\left(\mathbf{h}_{t}\right)\right)\right)\,,\nonumber 
\end{align}
where $\sigma\left(x\right)$ is the sigmoid function, $\tau\left(x\right)$
is the hyperbolic tangent, $\circ$ denotes an element-wise vector
multiply, and $\mathrm{GRU_{B}}\left(\cdot\right)$ is a regular (non-sparse but smaller)
GRU. The dual fully-connected ($\mathrm{dual\_fc}(\mathbf{x})$) layer is defined as
\begin{equation}
\mathrm{dual\_fc}(\mathbf{x})=\mathbf{a}_{1}\circ\tau\left(\mathbf{W}_{1}\mathbf{x}\right)+\mathbf{a}_{2}\circ\tau\left(\mathbf{W}_{2}\mathbf{x}\right)\,,\label{eq:dual_fc}
\end{equation}
where $\mathbf{W}_{1}$ and $\mathbf{W}_{2}$ are weight matrices
and $\mathbf{a}_{1}$ and $\mathbf{a}_{2}$ are scaling vectors.

The synthesized
excitation sample $e_{t}$ is obtained by sampling from the probability
distribution $p\left(e_{t}\right)$ after lowering the temperature, i.e., 
decreasing the entropy of the distribution,
of voiced frames as described in eq.~(7) of~\cite{lpcnet}. To 
reduce complexity, $\mathrm{GRU_{A}}$ uses sparse recurrent
matrices with non-zero blocks of size 16x1 to ensure efficient vectorization.
Because the hidden state update is more important than the reset and
update gates, we keep 20\% of the weights in $\mathbf{W}_{h}$, but
only 5\% of those in $\mathbf{W}_{r}$ and $\mathbf{W}_{u}$, for
an average of 10\%. If $N_A$ denotes the number of units in $\mathrm{GRU_{A}}$ 
the equivalent non-sparse number of units at a density $d$ is $\sqrt{dN_A^2 + N_A}$.

\begin{figure}
\centering{\includegraphics[width=1\columnwidth]{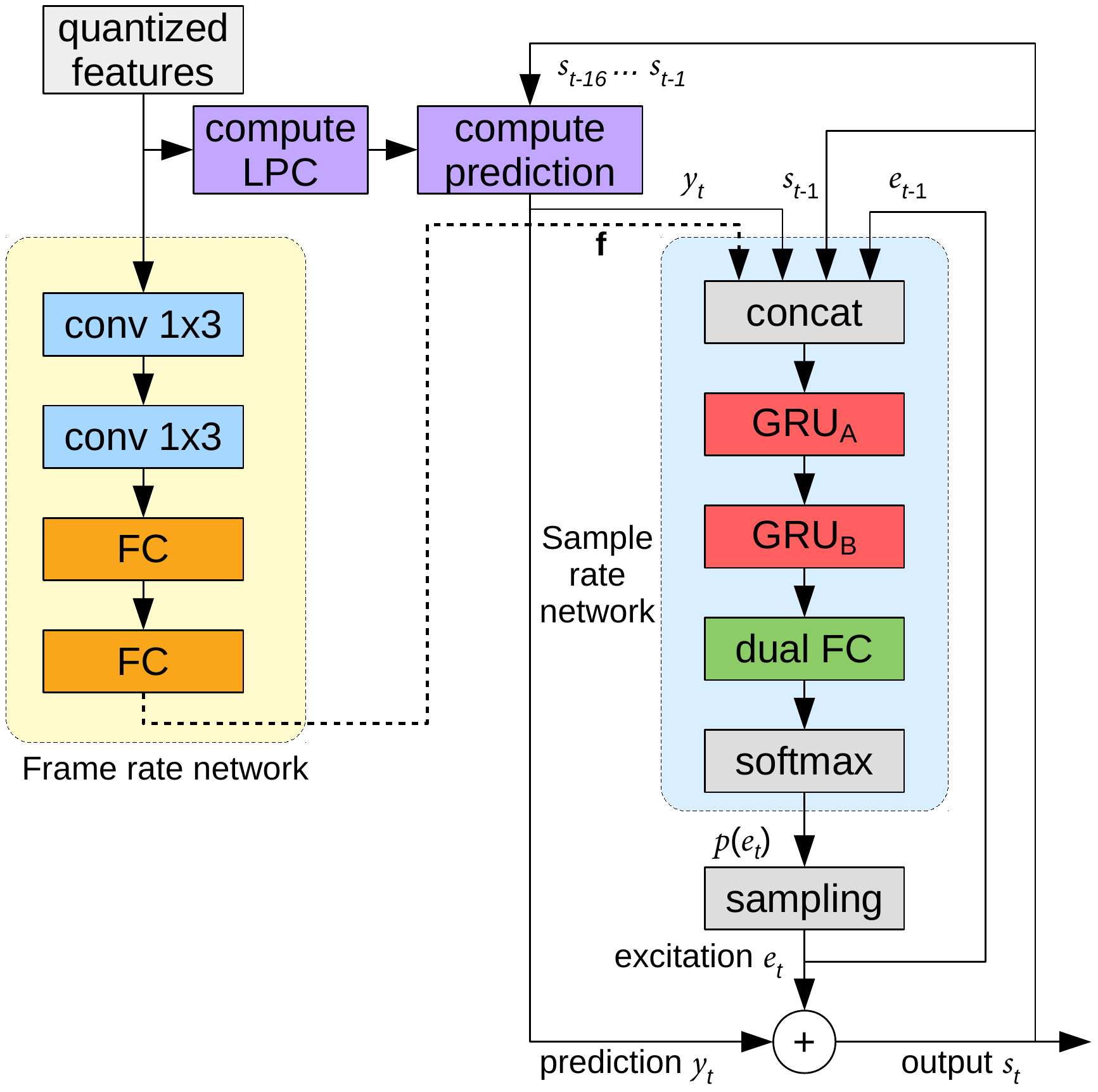}}
\caption{Overview of the LPCNet model. The frame rate network (yellow) operates
on frame-wise representing features and its output is held constant
through each frame for the sample rate network (blue). The \emph{compute
prediction} block applies linear prediction to predict the sample
at time $t$ from the previous samples. Conversions between $\mu$-law
and linear are omitted for clarity. The de-emphasis filter is applied
to the output $s_{t}$.\label{fig:Overview-of-LPCNet}}
\end{figure}

In this work we use a model with 384~units in $\mathrm{GRU_A}$
(equivalent to 122~non-sparse units) and 16~units for $\mathrm{GRU_B}$,
for a total of 72,000 weights in the sample rate network. This results in
a total complexity of 3~GFLOPS. The complexity was also measured on 
different CPU architectures. 
On x86, real-time synthesis requires 20\% of
one 2.4~GHz Broadwell core (5x~real-time).
On ARMv8 (with Neon intrinsics), real-time LPCNet synthesis on a
2.5~GHz Snapdragon~845 (Google Pixel~3) requires 68\% of one core
(1.47x~real-time). On the more recent 2.84~GHz Snapdragon~855 (Samsung
Galaxy~S10), real-time synthesis requires only 31\% of one core (3.2x~real-time).

\section{Conditioning Features and Training}
\label{sec:training}
%\tcr{Quarter page, half column.}
%(change wordings )
We mostly use the same set of conditioning features for both WaveNet and LPCNet.
Those features are extracted from the Opus bit stream and represent the
spectral shape and the pitch of the signal.
There are two ways to compute the spectral envelope of the decoded audio:
\begin{enumerate}
	\item Computing the spectrum on the decoded audio
	\item Converting the LPCs into a spectrum
\end{enumerate}

Both of those methods have significant drawbacks.
For low bit rates, method~1 suffers from the quantization noise in the residual.
On the other hand, while method~2 uses LPCs computed on clean speech, the fact
that the reference encoder jointly optimizes LPC and LTP causes the frequency
response of the LPC not to match the true spectrum of the signal.
Because of that, we use both methods and have two sets of spectral
features, from each of which we compute 18~cepstral coefficients.

The five pitch gains from the LTP are summed to produce a single pitch gain feature.
The sum represents the pitch gain for low frequencies, where it is
most relevant. The LTP pitch period is used directly. The two sets of cepstral coefficients
plus the two pitch parameters amount to a total of 38 features for LPCNet. The LPC
parameters are computed from the decoded cepstrum rather than from the LPC in the
Opus bit stream.

For LPCNet, best results are obtained when the whole network is trained on clean speech
and then only the frame rate network is adapted with the decoded speech features. As reported
in~\cite{jin2018fftnet}, adding noise to the network input signal can be beneficial to improve
robustness to the training-inference mismatch caused by teacher forcing. For that reason,
we add a small amount of Laplacian-distributed noise to the excitation inside the prediction loop,
as shown in Fig.~3 of~\cite{valin2019lpcnetcodec}.

For WaveNet, the feature set described above does not result in synthesis with acceptable speech quality.
Instead, only the cepstrum from method~2 is used and the model is trained directly
on the decoded speech features.

To make both models more robust to variations in the input, we augment the training
data. The signal level is varied over a 40 dB range and the frequency response is varied
according to eq.~(7) in~\cite{valin2017hybrid}).

\section{Evaluation}
\label{sec:evaluation}

The source code for the LPCNet model is available at \href{https://github.com/mozilla/LPCNet/}{https://github.com/mozilla/LPCNet/}
under a BSD license. The evaluation in this section is based on commit \texttt{2b64e3e}.
The conditioning features are produced using a slightly modified Opus decoder found
at commit \texttt{ec5bf39} of \href{https://github.com/xiph/opus/}{https://github.com/xiph/opus/}.

\subsection{Experimental Setup}
%\tcr{Quarter page, half column.}
The model is trained using four~hours of 16~kHz-sampled speech (wideband) from the NTT Multi-Lingual
Speech Database for Telephonometry (21~languages) \cite{ntt1994}. We
excluded all utterances from the speakers used in testing. Using the original
data, we generated 14~hours of augmented speech data as described
in Section~\ref{sec:training}.

We conducted a subjective listening test with a MUSHRA-inspired crowd-sourced methodology 
to evaluate the quality of neural synthesis of Opus parameters coded at 6~kb/s (average rate, since SILK is variable rate)
in wideband mode.
As an indication on the highest quality achievable with autoregressive neural synthesis,
albeit at intractable complexity and latency,
we included WaveNet synthesis. LPCNet synthesis represents a practically implementable 
system of today. We also compared with Opus~\cite{rfc6716} wideband (SILK mode) operating at 9~kb/s
VBR\footnote{The lowest bit rate for which the encoder defaults to wideband if signal bandwidth is not specifically set.}.
We omitted a 3.5kHz LP-filtered original as anchor, which is the standard anchor in MUSHRA~\cite{BS1534}, as it likely would not be considered as the lowest quality.
Instead, as a more appropriate low anchor we used Speex~\cite{valin2007speex} operating as a 4~kb/s wideband
vocoder (using the wideband quality setting at~0). The reference signal is sampled at 16~kHz.

\begin{figure}[t]
  \centering
  \centerline{\includegraphics[width=\columnwidth]{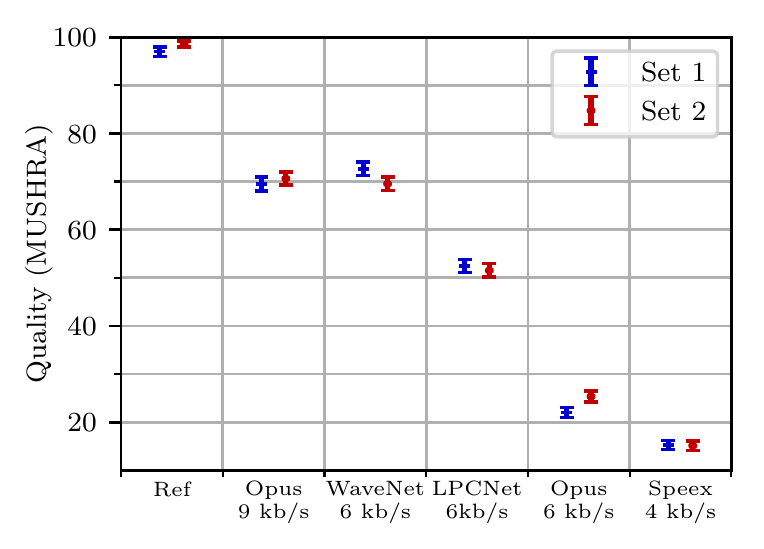}}
  \caption{Listening test results for sets~1 and~2. The error bars indicate a 95\% confidence interval.}
  \label{fig:results}
\end{figure}

In a first test (Set~1), we used eight~utterances from two~male and two~female
speakers. The utterances were from the NTT database used in training,
but all utterances from the selected speakers for the test were excluded
from the training set. As reported in~\cite{klejsa2019}, mismatches
between the training and testing databases can cause a significant
difference in the output quality. We measure that impact in a second
test (Set~2) on the same model, with eight~utterances (one~male and one~female
speaker) from the dataset used to create the Opus test vectors \cite{rfc8251}.
Each test included 100~listeners.

\subsection{Results}
We can see from the results in Fig.~\ref{fig:results} that even though Opus 
produces an unacceptable wideband quality at 6 kb/s, 
both WaveNet and LPCNet provide sufficient improvements
to make such a low rate usable.
WaveNet synthesis from a 6~kb/s bit stream has a quality level comparable
to Opus coded at 9~kb/s. LPCNet synthesis from the same bit stream yields a quality level 
that sits between Opus at 6~kb/s and 9~kb/s.

\section{Conclusion}
\label{sec:conclusion}
We have shown that neural synthesis can significantly improve the output of a
low bit rate Opus bit stream. Previous speech coding efforts using neural synthesis were 
based on pure parametric coding, here we expand the scope to address also a waveform matching 
coder. Furthermore, when using the LPCNet architecture,
real-time synthesis can be achieved even on a mobile device. This opens the door
to improving other existing speech coders, such as AMR-WB~\cite{bessette2002adaptive},
extending their life without breaking
compatibility. Also, in this work, synthesis is performed using only frequency
domain features, without directly using any temporal information from the decoded signal. In the future, even
better quality may be achievable by using temporal processing with the time-domain
decoded signal.

% -------------------------------------------------------------------------
% Either list references using the bibliography style file IEEEtran.bst
\bibliographystyle{IEEEtran}
\bibliography{opusnet}

\end{sloppy}
\end{document}